\documentclass[aps, reprint, 10pt, onecolumn, tightenlines, notitlepage, superscriptaddress, nofootinbib, preprintnumbers, floatfix]{revtex4}
\pdfoutput=1
\usepackage{epsfig,amsfonts,mathrsfs,amsmath,amssymb,graphicx,color,slashed,multirow}
\usepackage{amsmath,latexsym,amssymb,graphicx,slashed,hyperref,color,enumerate,url,etoolbox,cancel,gensymb}
\hypersetup{colorlinks,citecolor= nicegreen,linkcolor= nicered}
\definecolor{nicered}{rgb}{0.7,0.1,0.1}
\definecolor{nicegreen}{rgb}{0.1,0.5,0.1}
\newcommand{\boldgreek}[1]{{\mbox{\boldmath{$#1$}}}}
\newcommand{\bg}{\boldgreek}

\def\Fermilab{Theoretical Physics Department, Fermilab, P.O. Box 500, Batavia, IL 60510, USA}
\def\Northwestern{Department of Physics and Astronomy, Northwestern University, Evanston, IL 60208, USA}

\begin{document}

\title{Mono-Neutrino at DUNE: New Signals From Neutrinophilic Thermal Dark Matter}

\author{Kevin J. Kelly}
\affiliation{\Fermilab}
\author{Yue Zhang}
\affiliation{\Fermilab}
\affiliation{\Northwestern}

\date{\today}

\begin{abstract}
We introduce the {\it mono-neutrino} signal at neutrino detectors as a smoking gun of sub-GeV scale dark matter candidates that mainly interact with standard model neutrinos. In a mono-neutrino process, invisible particles, either dark matter themselves or the mediator particle, are radiated off a neutrino when it undergoes the charged-current weak interaction. The associated signals include a missing transverse momentum with respect to the incoming neutrino beam direction and the production of wrong-sign charged leptons. We demonstrate the potential leading role of the future DUNE experiment, using its proposed liquid and gas argon near detectors, in probing these new signals and  the thermal origins of neutrinophilic dark matter.
\end{abstract}

\preprint{FERMILAB-PUB-19-002-T}

\maketitle 

\section{Introduction}

The coming decade is about to witness impressive progress at the neutrino frontier, with new experiments to be built for further exploring the present unknowns in the neutrino sector, as well as any associated new physics. One in particular is the DUNE experiment~\cite{Acciarri:2016crz, Acciarri:2015uup, Acciarri:2016ooe}, which will be equipped with a high intensity beam and liquid argon neutrino detectors. With these new facilities, the door is open for exploring a broader range of physics beyond the standard model (SM), and hunting for the corresponding new signals. 

One of the exciting and well-motivated opportunities for DUNE as a multipurpose experiment is to probe the nature of dark matter.
While there is enormous evidence from cosmology suggesting the existence of dark matter~\cite{Jungman:1995df, Bertone:2004pz}, the burning question is how to look for it in our laboratories.
In terms of energy deposition, neutrino detectors typically have much higher energy thresholds than dedicated dark matter detectors and thus are not tailored for probing dark matter scattering as in the traditional direct detection picture~\cite{Goodman:1984dc}. Many variations to this conventional picture have been proposed, such as searching for a more energetic component of dark matter (or dark radiation) in our galaxy which arises from astrophysical processes~\cite{Agashe:2014yua, Chatterjee:2018mej, Bringmann:2018cvk, Ema:2018bih}, or a man-made dark matter beam created via fixed-target collisions which then strikes the detector to scatter elastically~\cite{Batell:2009di, Coloma:2015pih, Aguilar-Arevalo:2017mqx, Aguilar-Arevalo:2018wea} or causes charged lepton pair creations~\cite{deGouvea:2018cfv}. It has also been envisioned that a fraction of the halo dark matter may live in a metastable bound state and release much more than its kinetic energy once it interacts~\cite{Grossman:2017qzw}. Note that, in all scenarios mentioned above, dark matter particles originate  outside the neutrino detectors. Despite using a neutrino facility for detection, the key interactions between dark matter and the SM sector here do not involve neutrinos.

New dark-matter-neutrino interactions can leave important imprint in various aspects of cosmology, such as the cosmic microwave background (CMB)~\cite{Serra:2009uu, Wilkinson:2014ksa, DiValentino:2017oaw}, and the structure formation from large to small scales~\cite{Mangano:2006mp, Hannestad:2013ana, Dasgupta:2013zpn, Bringmann:2013vra, Cherry:2014xra, Bertoni:2014mva, Binder:2016pnr, Davoudiasl:2018hjw}. 
Experimentally, neutrinos from dark matter annihilation at the center of the galaxy or the sun have long been proposed and searched for using neutrino telescopes~\cite{Desai:2004pq, Choi:2015ara, Aartsen:2015xej, Adrian-Martinez:2015wey, Srednicki:1986vj, Zentner:2009is, Chen:2011vda, Gao:2011bq, Batell:2017rol, Campo:2017nwh, Berryman:2017twh, McKeen:2018pbb}.
The attenuation effect in the flux of high energy neutrinos from distant astrophysical sources due to the travel through the cosmic dark matter background has also been explored in Refs.~\cite{Ng:2014pca, Cherry:2016jol, Arguelles:2016zvm, Capozzi:2017auw, Kelly:2018tyg}.
The direct impact of dark-matter-neutrino interactions on the observed neutrino oscillation probabilities~\cite{Berlin:2016woy, Krnjaic:2017zlz, Brdar:2017kbt} has been recently investigated in the context of fuzzy dark matter models~\cite{Hu:2000ke, Hui:2016ltb, Davoudiasl:2017jke}. These existing works represent the wide range of dark matter theory space and the corresponding signals that can be searched for using multipurpose neutrino detectors.

In this work, we present a new type of dark matter signal at neutrino experiments from neutrinophilic dark matter candidates, which mainly interact with neutrinos rather than other SM particles. We introduce the {\it mono-neutrino} process where invisible particles, either dark matter themselves or the mediator to dark matter, are radiated away while the neutrino undergoes a charged-current (CC) interaction and gets detected. The final states of such a process are similar to the normal neutrino CC interaction, but there is an imbalance of the final state momenta transverse to the neutrino beam direction. This can be viewed as an expansion of the mono-$X$ search concept for dark matter at high energy collider experiments~\cite{Abercrombie:2015wmb}, where $X$ is now a neutrino and the search could only be performed at neutrino detectors. 
The next-generation liquid argon detectors seem to be the best places for this search thanks to their unprecedented particle identification and energy resolution capabilities.
We will examine several neutrinophilic dark matter models, and then identify the parameter space where DUNE using its near detectors will be at the frontier of probing the thermal origin of these dark matter candidates. 

This paper is organized as follows. In section~\ref{sec:benchmark} and \ref{sec:relics}, we present the benchmark models for neutrinophilic dark matter and calculate their thermal relic abundance through the annihilation into neutrinos, respectively. We present the mono-neutrino process in section~\ref{sec:mononu@DUNE}, discuss the characteristic signatures, and then explore the prospects of probing the associated signals at the future DUNE near detectors. In the models under consideration, the mono-neutrino signal is always accompanied with the ``wrong-sign'' charged lepton production. We point out that the DUNE reach can be further improved with a charge identification capability. We wrap up after commenting on a few other existing constraints in section~\ref{sec:otherbounds}.

\section{Benchmark models}\label{sec:benchmark}

To set the stage, we introduce a complex scalar particle $\phi$ which carries lepton number (or the $B-L$ quantum number) equal to $-2$ ($2$) and couples to SM neutrinos through the Weinberg operator portal
\begin{equation}\label{Leff}
\mathcal{L}_{\rm portal} = \frac{(L_\alpha H)(L_\beta H)}{\Lambda_{\alpha\beta}^2} \phi + {\rm h.c.} \to \frac{1}{2} \sum_{\alpha,\beta=e,\mu,\tau} \lambda_{\alpha\beta} \nu_\alpha \nu_\beta \phi + {\rm h.c.} \ .
\end{equation}
After electroweak symmetry breaking, the Higgs vacuum expectation value (VEV) projects out only neutrinos in the unitary gauge, as shown in the second step, making $\phi$ neutrinophilic.
The dimensionless coupling $\lambda_{\alpha\beta}$ is defined as $v^2/\Lambda_{\alpha\beta}^2$ with $v=246\,$GeV.

The above $\phi$ coupling looks similar to that of a Majoron~\cite{Gelmini:1980re, Barger:1981vd}, the goldstone boson (real scalar) from spontaneously broken lepton number (or $B-L$)  symmetry. If such a symmetry breaking is responsible for generating the neutrino mass (Majorana in this case), the Majoron coupling will be equal to $\lambda = i m_\nu/f$, where $f$ is the symmetry breaking scale. In this work, we are interested in sizable couplings $\lambda\sim\mathcal{O}(1)$, which in turn imply very low $f \sim m_\nu$. Therefore, for processes occurring at much higher energies, {\it e.g.}, at neutrino experiments, one should observe the effect of symmetry restoration. In other words, we must consider neutrinos coupling to the whole complex scalar field containing both the Majoron as well as the radial excitations along the $f$ direction. The corresponding Lagrangian is identical to that of $\phi$ in Eq.~(\ref{Leff}).

Alternatively, one could assume that the $\phi$ field does not develop a VEV.
As argued in~\cite{Berryman:2018ogk}, introducing a $\phi$-like scalar is the easiest way of restoring the $B-L$ as an exact global symmetry of the standard model effective theory. Eq.~(\ref{Leff}) is the lowest dimensional non-renormalizable operator for $\phi$ to couple to standard model particles.
In such a context, the neutrino mass has to be Dirac, and one must introduce at least two right-handed neutrinos (denoted as $\nu_R$), with Yukawa coupling to the SM lepton doublets. A marginal operator, $\lambda'\nu_R\nu_R\phi$, is also allowed.
In the presence of both $\lambda$ and $\lambda'$ couplings, the right-handed neutrinos could be produced through the $\phi$ exchange in the early universe.
To avoid an excessive contribution to the neutrino degrees of freedom ($\Delta N_\mathrm{eff}$), which is tightly constrained by the CMB observations~\cite{Aghanim:2018eyx}, the product of couplings, $\lambda \lambda'$ must be tiny, $\lesssim 10^{-9}$ for GeV scale $m_\phi$~\cite{Berryman:2018ogk}. 

Within the parameter space of interest to this work,
the new phenomena we are going to explore based on Eq.~(\ref{Leff})
do not depend on the details of neutrino mass generation, whether Dirac or Majorana.

We continue by assuming $B-L$ conservation at the Lagrangian level, and 
further consider the above neutrinophilic $\phi$ boson as the portal to dark matter. We will examine a few simple models where the dark 
matter is stabilized by various symmetries, $\mathbb{Z}_2$, $\mathbb{Z}_3$ and $U(1)$.
\begin{equation}
\begin{split}
{\it Model\ IA}:\hspace{1cm}&\mathcal{L} = \frac{1}{2} y_{IA} \chi^2 \phi + {\rm h.c.} \\
{\it Model\ IB}:\hspace{1cm}&\mathcal{L} = \frac{1}{2} y_{IB} \bar\chi^c \chi \phi + {\rm h.c.} \\
{\it Model\ II}:\hspace{1cm}&\mathcal{L} = \frac{1}{6} y_{II} \chi^3 \phi + {\rm h.c.}\\
{\it Model\ III}:\hspace{1cm}&\mathcal{L} = y_{III} \bar \chi_1 \chi_2 \phi  + {\rm h.c.}\\
\end{split}
\end{equation}
In model IA, the dark matter $\chi$ is a complex scalar carrying lepton number $+1$ and is stabilized by a $\mathbb{Z}_2$ symmetry. 
The coupling $y_{IA}$ in this model has mass dimension 1.
Model IB is similar to IA but with $\chi$ being a Dirac fermion and $\chi^c$ its charge-conjugation field. In this case, $\chi$ has the same quantum numbers as the right-handed
neutrino and therefore its stability is not automatic.
An {\it ad hoc} $\mathbb{Z}_2$ symmetry under which $\chi$ is odd must be imposed by hand.
We include this case in our study for the sake of completeness. 
In model II, the dark matter carries lepton number $+2/3$ and is stabilized by a $\mathbb{Z}_3$ symmetry.
In this case, $\chi$ must be a complex scalar.
Model III slightly goes beyond the minimality and contains two components of dark matter, $\chi_1, \chi_2$, which carry lepton numbers $\mathtt{q}-2$ and $\mathtt{q}$ respectively. We will assume $\chi_1, \chi_2$ are Dirac fermions. For generic values of $\mathtt{q}$ their couplings to SM particles are forbidden, thus the lighter of the two (assumed to be $\chi_1$) can be the dark matter candidate. In this case, the stabilization symmetry is a $U(1)$.

It is worth pointing out that the $\mathbb{Z}_2$, $\mathbb{Z}_3$ and $U(1)$ symmetries in the above models (with the exception of model IB) and the stability of dark matter 
are the natural consequences of the assumed $B-L$ conservation. The $B-L$ quantum number assignment forbids any renormalizable interaction for the dark matter candidate to decay into standard model particles.
For example, the $\mathbb{Z}_2$ symmetry in model IA is the analogue of $R$-parity in $B-L$ conserving supersymmetric theories (where $\chi$ is like a right-handed sneutrino).

The above models will serve as the benchmarks for our study. In the following, we will first derive the favored regions of parameter space in each model 
that give the correct dark matter relic density via the thermal freeze out mechanism. Afterwards, we investigate the prospects of using the DUNE near detector(s) to test these well-motivated thermal targets.

\section{Dark matter thermal relic density}\label{sec:relics}

\begin{figure}[h!]
\centerline{\includegraphics[width=0.618\textwidth]{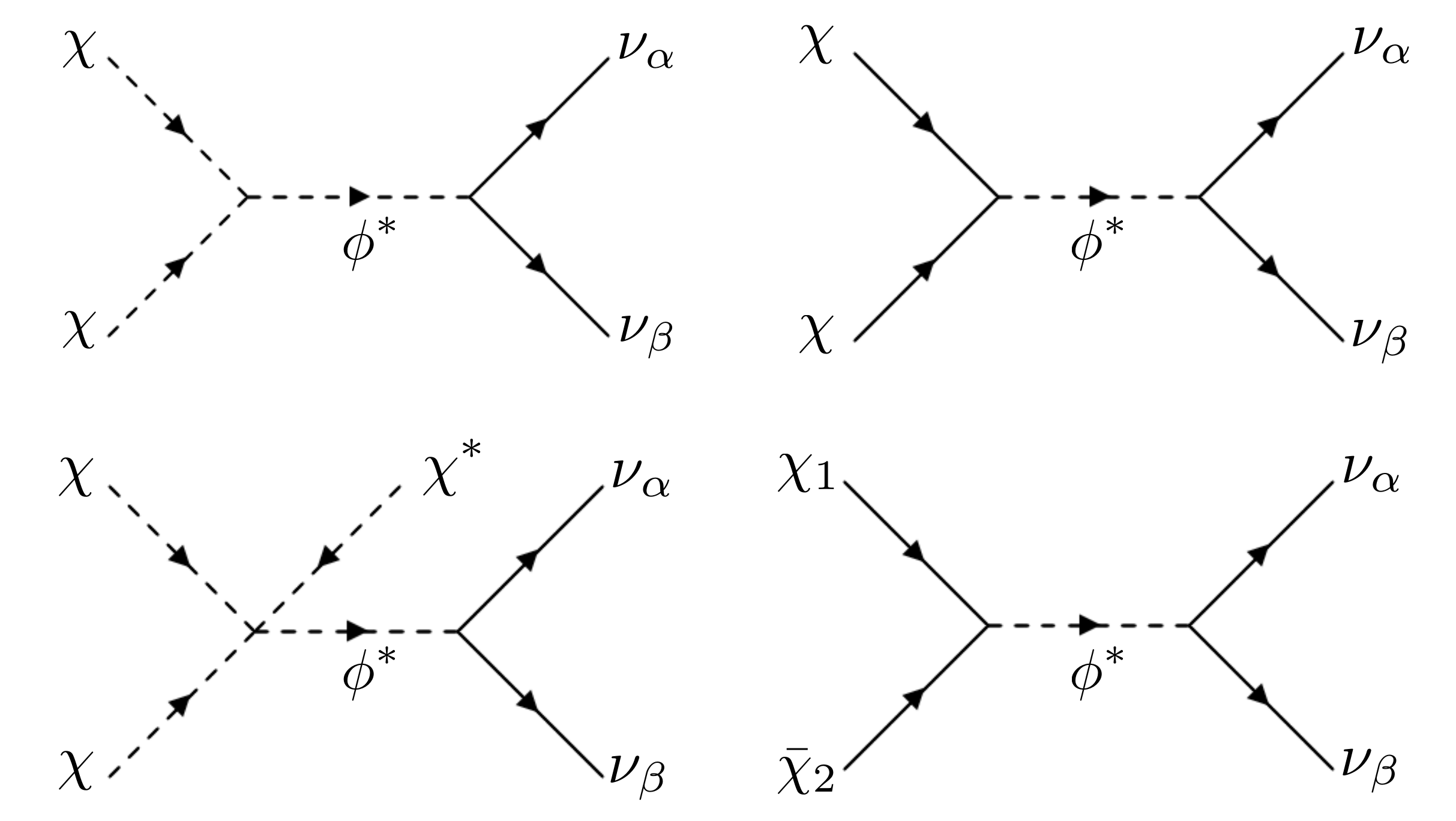}}
\caption{Annihilation channels for dark matter to freeze out in model IA (upper left), IB (upper right), II (lower left) and III (lower right).
Time flows from left to right. Arrows represent the direction of lepton number flows.}\label{Fig:Ann}
\end{figure}

We consider the thermal relic abundance of dark matter described in the above models. 
The relevant annihilation processes are in depicted in Fig.~\ref{Fig:Ann}, where the dark matter freezes out by 
annihilating directly to SM neutrinos via an off-shell mediator $\phi$.
We will focus sub-GeV dark matter and mediator mass scales, which makes them kinematically accessible in neutrino experiments. 
Throughout this study, we assume the mediator $\phi$ to be heavier than the dark matter $\chi$ or $\chi_{1,2}$.\footnote[2]{If $m_\chi > m_\phi$, the freeze-out of $\chi$ could occur through the annihilation into $\phi$'s instead of neutrinos. The subsequent decay $\phi \to \nu_\alpha \nu_\beta$ is allowed to happen on a much longer time scale. In such a secluded scenario, there is no direct connection between the $\phi-\nu$ coupling $\lambda$, which could be probed at DUNE (this work), and the thermal relic abundance of dark matter $\chi$.
For this reason, we will restrict ourselves to the heavier mediator case ($m_\phi > m_\chi$).}
The annihilation cross sections are thus proportional to the square of $\lambda_{\alpha\beta}$ couplings which, as we show later on, could manifest themselves at DUNE. 

\begin{figure}[t]
\centerline{\includegraphics[width=0.7\textwidth]{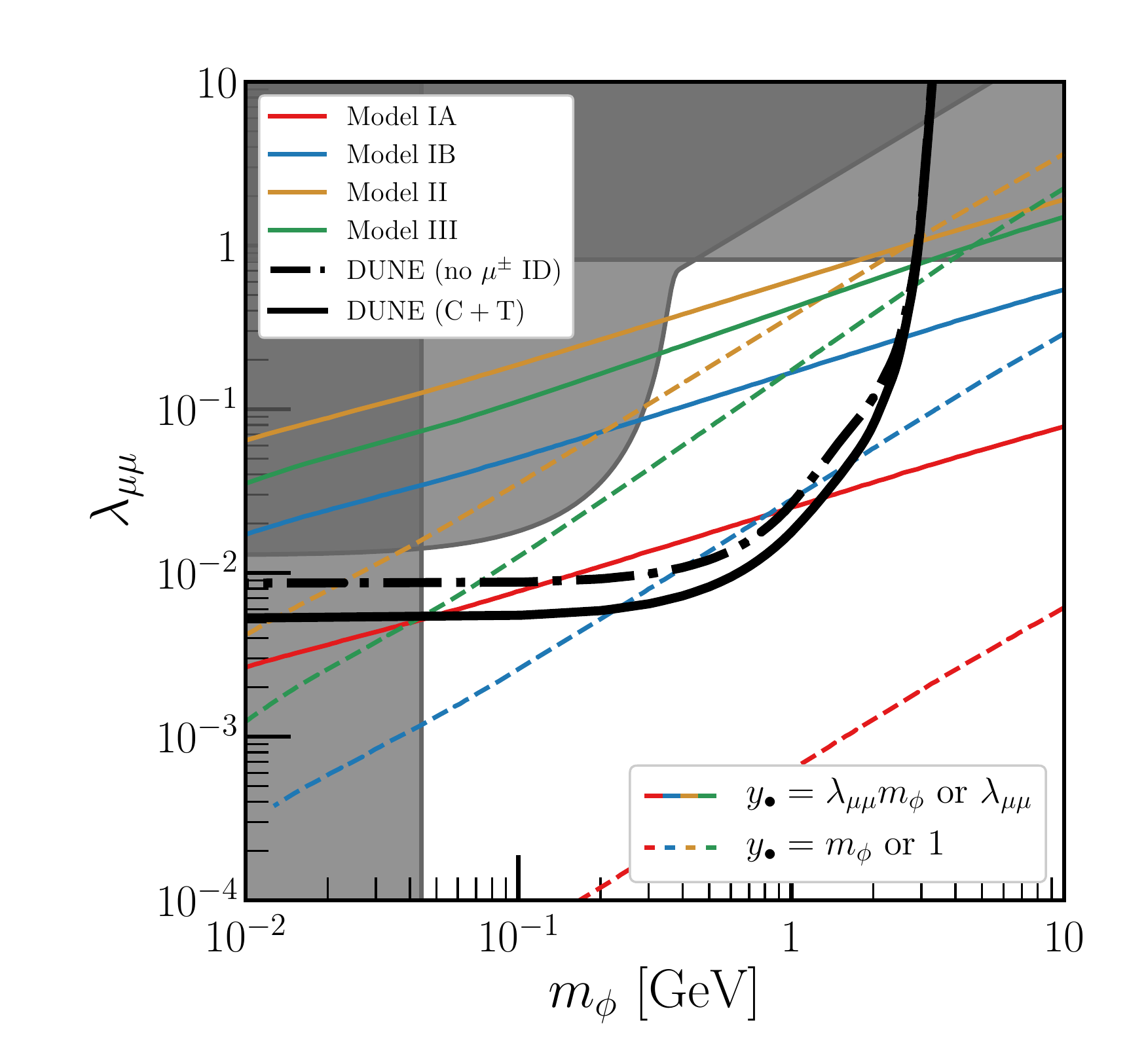}}
\caption{
Expected reach of mono-neutrino search at DUNE versus the theory landscape of neutrinophilic dark matter with thermal relic density:
{\sl Model IA}: solid (dashed) red curve corresponds to $y_{IA}/m_\phi=\lambda_{\mu\mu}$ ($y_{IA}/m_\phi=1.0$);
{\sl Model IB}:  solid (dashed) blue curve corresponds to $y_{IB}=\lambda_{\mu\mu}$ ($y_{IB}=1.0$);
{\sl Model II}: solid (dashed) yellow curve corresponds to $y_{II}=\lambda_{\mu\mu}$ ($y_{II}=1.0$);
{\sl Model III}: solid (dashed) green curve corresponds to $y_{III}=\lambda_{\mu\mu}$ ($y_{III}=1.0$) and $\Delta=0.2$.
We assume $m_\chi=m_\phi/10$ for all the curves.
Also shown are the present and future experimental constraints.
The gray shaded regions are already ruled out by the charged kaon semi-leptonic decays (upper-left region), the Higgs boson invisible decay (large $\lambda_{\mu\mu}$ region) 
and the CMB (small $m_\phi$ region) measurements, as will be discussed in section~\ref{sec:otherbounds}.
The solid and dot-dashed thick black curves show the DUNE mono-neutrino reach with and without taking into account of charge identification,
as will be discussed in detail in section~\ref{sec:mononu@DUNE}.
}\label{moneyplot}
\end{figure}

In each model, the dark matter annihilation cross section is calculated in the low relative velocity $v_{\rm rel}$ limit, keeping the leading term in the $v_{\rm rel}$ expansion.

\smallskip
{\it Model IA}: For scalar $\chi$, the annihilation is $S$-wave,
\begin{equation}
\sigma v_{\rm rel} (\chi \chi \to \nu_\alpha\nu_\beta) = \frac{|\lambda_{\alpha\beta} y_{IA}|^2}{8\pi(4m_\chi^2-m_\phi^2)^2(1 + \delta_{\alpha\beta})} \ .
\end{equation}

\smallskip
{\it Model IB}: For fermionic $\chi$, the annihilation is $P$-wave,
\begin{equation}
\sigma v_{\rm rel} (\chi \chi \to \nu_\alpha\nu_\beta) = \frac{|\lambda_{\alpha\beta} y_{IB}|^2 m_\chi^2 v_{\rm rel}^2}{16\pi(4m_\chi^2 - m_\phi^2)^2(1 + \delta_{\alpha\beta})} \ .
\end{equation}

For the dark matter relic density, we follow the semi-analytical approach in Refs.~\cite{Griest:1990kh, Kolb:1990vq},
\begin{equation}\label{KTbook}
\begin{split}
\Omega_{\rm DM} h^2 &= \frac{2.1\times 10^{9} \,{\rm GeV}^{-1} \sqrt{g_*} (n+1) x_f^{n+1}}{g_{*S} M_{pl} \sigma_0} \ , \\
x_f &\simeq \log \frac{0.038 g_\chi c(c+2) M_{pl} m_\chi \sigma_0}{\sqrt{g_*}} \ ,
\end{split}
\end{equation}
where $g_\chi=2$. For $S$-wave annihilation, $n=0$ and $\sigma_0=\sigma v_{\rm rel}$; while for $P$-wave annihilation, $n=1$ and $\sigma_0=6 \sigma /v_{\rm rel}$. We set $c(c+2)=n+1$ following the prescription in Ref.~\cite{Kolb:1990vq}. $\Omega_{\rm DM} h^2$ includes contributions from both $\chi$ and $\chi^*$ ($\chi^c$).

\smallskip
{\it Model II}: The dark matter in this model freezes out via semi-annihilation~\cite{DEramo:2010keq},
\begin{equation}
\sigma v_{\rm rel} (\chi \chi \to \chi^* \nu_\alpha\nu_\beta) = \frac{|\lambda_{\alpha\beta} y_{II}|^2}{2048\pi^3 m_\chi^2 (1+\delta_{\alpha\beta})}  \int_0^1 dz \frac{z\sqrt{(1-z)(9-z)}}{(m_\phi^2/m_\chi^2 - z)^2} \ .
\end{equation}
This integral can be performed analytically, but we will not show the lengthy full result except for the large $m_\phi$ limit, 
\begin{equation}
\sigma v_{\rm rel} (\chi \chi \to \chi^* \nu_\alpha\nu_\beta) \simeq \frac{(57-80\ln2) |\lambda_{\alpha\beta} y_{II}|^2 m_\chi^2}{4096\pi^3 m_\phi^4 (1+\delta_{\alpha\beta})} \ .
\end{equation}
This is an $S$-wave semi-annihilation.
The dark matter relic density can be calculated by generalizing Eq.~(\ref{KTbook}), only with the change $c(c+2)\to c(c+1)$. However, we still set $c(c+1)=n+1$ in the relic density calculation.

\smallskip
{\it Model III}: In this model, the way to freeze out dark matter in early universe is via coannihilation of $\chi_1, \chi_2$, whose cross section is
\begin{equation}
\sigma v_{\rm rel} (\chi_1 \bar\chi_2 \to \nu_\alpha\nu_\beta) = \frac{|\lambda_{\alpha\beta} y_{III}|^2 m_{\chi_1}^2(2+\Delta)^2 v_{\rm rel}^2}{64\pi[ m_{\chi_1}^2(2+\Delta)^2 - m_\phi^2 ]^2(1 + \delta_{\alpha\beta})} \ ,
\end{equation}
where $\Delta\equiv (m_2-m_1)/m_1$.
Like model IB, this is again a $P$-wave annihilation. The dark matter relic density in this case can be calculated using Eq.~(\ref{KTbook}) but with the replacement~\cite{Griest:1990kh}
\begin{equation}
\begin{split}
g_\chi &\to g_{\rm eff} = g_\chi \left[ 1 + (1+\Delta)^{3/2} e^{-x_f \Delta} \right] \ , \\
\sigma_0 &\to \frac{6 \sigma }{v_{\rm rel}} \times 2 (1+\Delta)^{3/2} e^{-x_f \Delta}  \ .
\end{split}
\end{equation}

Fig.~\ref{moneyplot} shows parameter space where the dark matter in the above models can obtain the observed relic density. 
For each model, we choose two benchmark values of the $\chi$-$\phi$ coupling, democratic and hierarchical. 
Namely, $\phi$ having similar branching ratios decaying into $\chi$'s or $\nu$'s, and $\phi$ mainly decaying into $\chi$'s, respectively.
For model IA, the solid (dashed) curve corresponds to $y_{IA}/m_\phi=\lambda_{\mu\mu}$ ($y_{IA}/m_\phi=1$).
For models IB, II and III, the solid curves stand for the case where $\phi$ couples with equal strength to dark matter and neutrinos ($y_\bullet=\lambda_{\mu\mu}$ where $\bullet=IB, II, III$), 
whereas the dashed curves stand for the case of hierarchical couplings with $y_\bullet =1$.
In model III, we set the $\chi_1$-$\chi_2$ mass difference to be $\Delta=0.2$.
We also set $m_\chi=m_\phi/10$.
These benchmark parameters are chosen in order to give an idea of the range of thermal dark matter targets for experimental searches -- the main focus of the upcoming sections.

\section{Mono-neutrino at DUNE near detectors}\label{sec:mononu@DUNE}

In this section, we propose the mono-neutrino channel as the new signal from $\phi$ as the mediator to dark matter with a sub-GeV mass, and explore the exciting opportunity of probing it at accelerator neutrino experiments, in particular the near-future DUNE experiment.
Such a signal is commonly predicted by the above benchmark models.
We consider neutrino near detectors which are located near the fixed-target collisions and see the most intense flux of neutrinos.
Because the accelerator neutrinos are predominantly made of the muon flavor at production, and charged muons are easiest to identify experimentally, we will focus on the $\lambda_{\mu\mu}$ coupling of $\phi$.

\subsection{The mono-neutrino signal and background}

\begin{figure}[t]
\centerline{\includegraphics[width=0.4\textwidth]{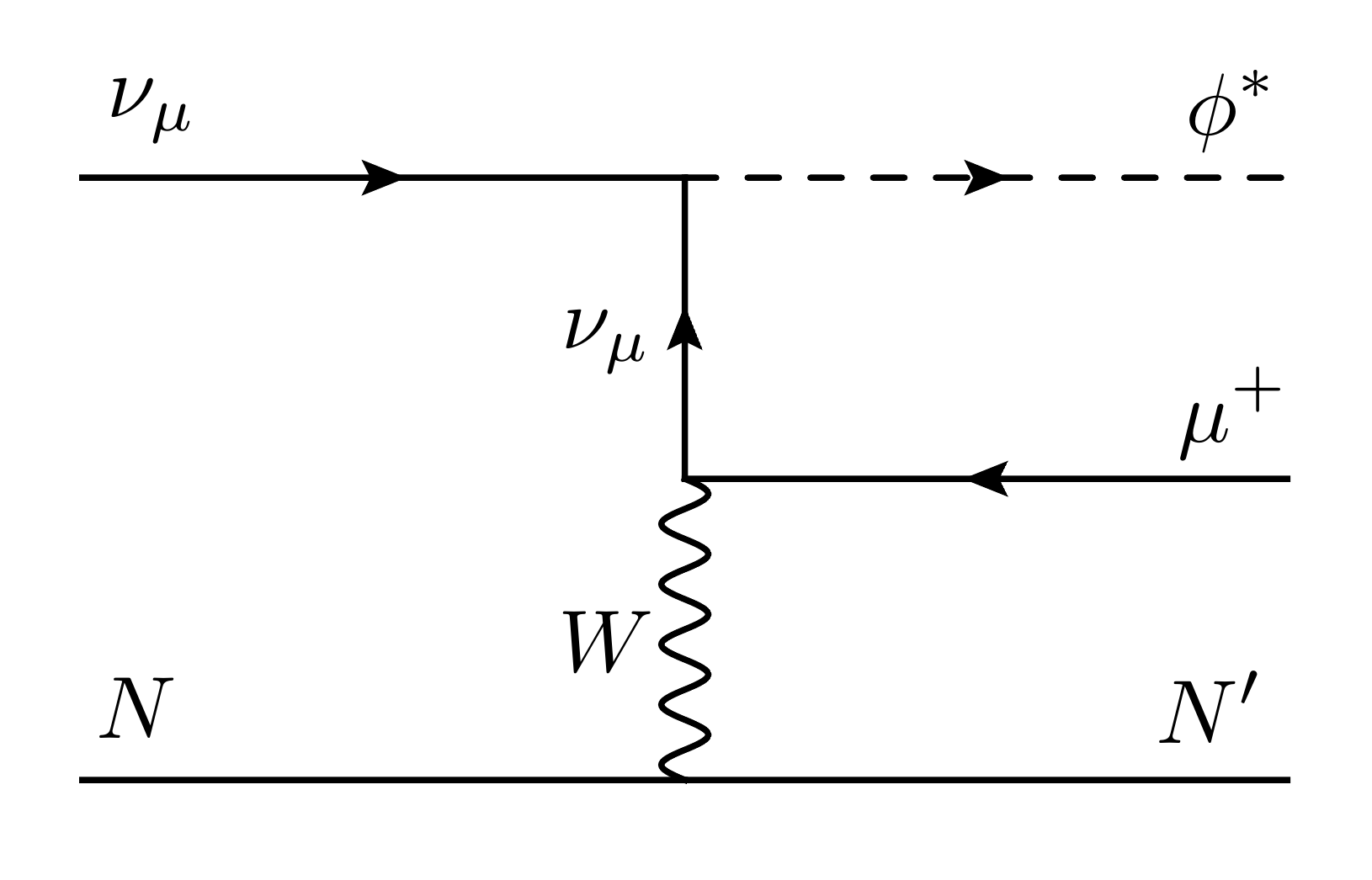}}
\caption{Neutrino beamstrahlung process that appears as a mono-neutrino event at DUNE. The $\phi$ particle is radiated off the incoming neutrino 
when a CC process occurs, followed by its invisible decay into dark matter or neutrinos. 
Time flows from left to right. Arrows represent the direction of lepton number flow.}\label{Fig:Brem}
\end{figure}

First, imagine an accelerator neutrino experiment whose beam is running in the neutrino mode, the key process of interest to us is
\begin{equation}\label{mononu}
\nu_\mu + N \to \mu^+ + N' + \phi^* 
\end{equation}
as depicted in Fig.~\ref{Fig:Brem}, where an on-shell $\phi^*$ particle is radiated off the initial state muon neutrino before it undergoes the SM CC weak interaction. 
Such neutrino beamstrahlung process has been proposed in~\cite{Berryman:2018ogk}.
In the models considered in this work, the $\phi^*$ particle will subsequently decay into dark matter or neutrinos and appear as missing energy. 
Because $\phi^*$ takes away two units of lepton number, a $\mu^+$ particle is created in (\ref{mononu}), accompanied with
a momentum transfer to the target nucleus, leading to nuclear activities (labelled by $N'$).
We will restrict our study to those events where the final state $N'$ remains as a nucleon, as additional hadronic activity ({\it e.g.}, deeply-inelastic scattering events) tends to bring more uncertainties to the final state momentum measurement.
If the neutrino detector in which the above process occurs is able to
identify the outgoing muon and the recoiling nucleon, and measure their energy and momenta precisely,
it will discover an imbalance among the final state momenta of these visible particles in the directions orthogonal to the incoming neutrino beam. 
The missing transverse momentum, defined as
\begin{equation}
\cancel{p}_T \equiv \left| \sum_{i=\mathrm{visible}} \left(\vec{p}_T\right)_i \right| \ ,
\end{equation}
could serve as the smoking gun for neutrinophilic dark matter through the $\phi$ portal. 
This is the {\it mono-neutrino} signal we shall explore in detail.

The main backgrounds for such signal are the SM charged current quasi-elastic (CCQE) events, 
\begin{eqnarray}
&& \nu_\mu + N \to \mu^- + N' \label{Bkg1} \\
&& \bar \nu_\mu + N \to \mu^+ + N' \label{Bkg2}
\end{eqnarray}
There are no invisible final state particles in these processes. However, with a realistic detector, the uncertainties in the final state momentum reconstruction (more uncertain on the nucleon side) could result in a nonzero $\cancel{p}_T$ measurement, in which case they fake the mono-neutrino signal.

Although we are considering a neutrino beam in the above example, there is always an anti-neutrino contamination in the beam thus the process in (\ref{Bkg2}) also needs to be taken into account.
Given that the antineutrino flux is only a few percent of that of neutrinos, the rate for (\ref{Bkg2}) is subdominant to (\ref{Bkg1}). 
However, because (\ref{Bkg2}) produces a muon with the same sign as that in our signal process (\ref{mononu}), 
it contributes as a more irreducible background even for detectors with a charge identification capability.

Another potential background worth worrying about arises from the neutral current process 
\begin{eqnarray}\label{NCpion}
\nu + N \to \nu + N' + \pi^\pm 
\end{eqnarray}
with the final state $\pi^\pm$ misidentified as a charged muon. 
In this case, such a process can fake our mono-neutrino signal because the final-state neutrino takes away a missing transverse momentum. The number of these events at the DUNE near detector is expected to be around half the number of CCQE events~\cite{Acciarri:2015uup}. There are, however, several useful handles for reducing this contribution.

First, such a process dominantly proceeds via an intermediate $\Delta$-baryon resonance, whose contribution can be efficiently suppressed with a cut on the  
$N' \pi^\pm$ invariant mass vetoing the window around $\Delta$ mass.
Second, the discrimination between the $\pi^\pm$ and $\mu^\pm$ tracks is possible using the $dE/dx$ observable~\cite{DUNENDPhysics} (especially in a gaseous argon detector as discussed below).
Third, because $\pi^\pm$ interacts strongly, it tends to shower more, producing a number of ancillary tracks as it travels through the liquid argon~\cite{pionAr}.
Fourth, the charged pion once produced could either get absorbed by an argon nucleus, or decay weakly into $\mu^\pm$.
The former possibility allows (\ref{NCpion}) to fake our signal, while the latter features a track with a kink (when $\pi^\pm$ and $\mu^\pm$ travel in different directions), which is relatively easy to identify.
The ratio of these two possibilities is in principle calculable. 
Therefore, measuring the rate of the tracks with kinks will help to have control of the background from absorbed pions.
Last but not the least, the Michel electrons from $\mu^\pm$ decay could serve as an important handle on identifying our signal process, which is absent in the $\pi^\pm$ decay. For all the above considerations, we decide to neglect the $\nu_\mu N \to \nu_\mu N' \pi^\pm$ background process in the following simulations and sensitivity estimates.

Atmospheric neutrinos with CC interactions could, in principle, fake our desired signal as well, since the incoming neutrino direction is unknown. However, the rate of atmospheric neutrino events at the DUNE near detector is expected to be small (less than 10 per year), and additionally, timing information with the DUNE beam spills would allow these events to be further reduced, since our desired signal arrives in time with the neutrino beam.

\subsection{Features of mono-neutrino events at DUNE near detectors}

\begin{figure}[t]
\centerline{\includegraphics[width=0.7\textwidth]{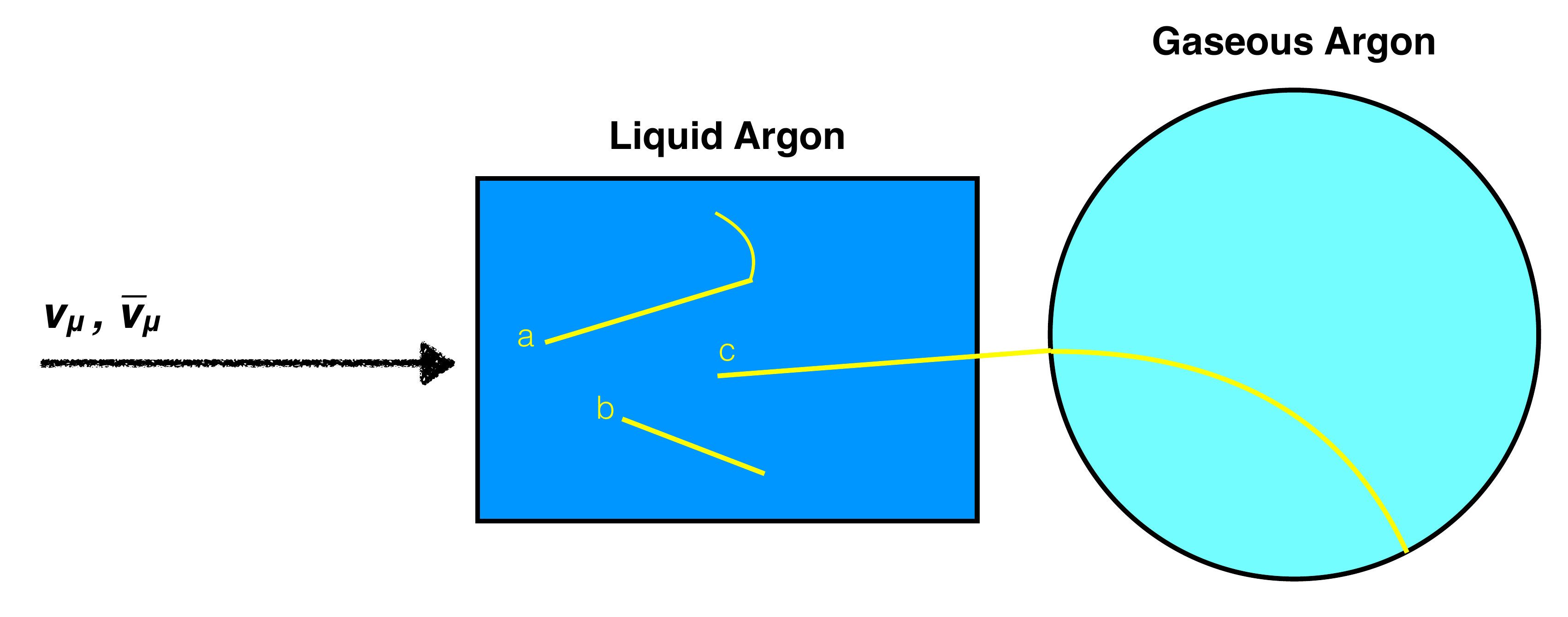}}
\caption{A schematic plot of the DUNE near detector setup, with liquid and magnetized gaseous argon detectors located downstream the neutrino beam.}\label{fig:schematic}
\end{figure}

Clearly, in order to suppress the background discussed above, the most useful neutrino detectors need to have excellent energy/momentum resolution and charge discrimination of final state leptons. This leads us to consider the future DUNE near detector setup, which according to the present plans~\cite{DUNENDComplex, DUNENDPhysics}, includes a liquid argon (LAr) time projection chamber (TPC)~\cite{DUNENDLAr}, a high-pressure gaseous argon (HPg) TPC~\cite{Martin-Albo:2016tfh, DUNENDGAr}, a 3D projection scintillator tracker~\cite{DUNENDScin}, as well as the DUNE PRISM concept of designing movable detectors~\cite{DUNEPRISM}.
Among them, of the most interest to this study are the LAr and HPg TPCs, which will be placed downstream the neutrino beam from the fixed target collision hall.
A schematic plot of the detector setup is shown by Fig.~\ref{fig:schematic}.
For a mono-neutrino process initiated in the LAr TPC, the final state muon and nuclear recoil leave distinct tracks, 
based on which their energies and momenta could be measured~\cite{Acciarri:2015uup}.
If the muon is energetic enough to travel into the HPg TPC, the sign of its charge can be easily identified using a magnetic field, in together with its energy and momentum.
Although the LAr TPC will not be magnetized, it still retains certain charge identification ability by utilizing the Michel electron from muon decay, which will be discussed in more detail below.

Our goal here is to estimate the reach of these DUNE near detectors in search for the mono-neutrino signal. We will focus on the processes initiated in the LAr TPC rather than the HPg one because the former has a much higher target mass. Based on the expected neutrino beam flux provided in~\cite{Acciarri:2015uup} and an effective LAr target mass equal to 75 ton~\cite{DUNENDComplex}, the SM CCQE background is estimated to occur about 20 million times every year~\cite{Acciarri:2015uup}.
To derive the final state phase space distributions for the signal and background, we simulate the corresponding processes at nucleon level using {\tt MadGraph}~\cite{Alwall:2011uj}, assuming $N=p, N'=n$ in (\ref{mononu}) and (\ref{Bkg2}) and $N=n, N'=p$ in (\ref{Bkg1}).
We impose an energy smearing of $40\%/\sqrt{E\ [\rm GeV]}$ for the final state neutron, $20\%/\sqrt{E\ [\rm GeV]}$ for proton, and $3\%/\sqrt{E\ [\rm GeV]}$ for muon, on event by event basis. These smearings account for the errors in final state momentum reconstructions~\cite{Acciarri:2015uup}.
We first generate the events for fixed neutrino energy and then convolve them with the incoming neutrino beam energy spectrum.

\begin{figure}[t]
\centerline{\includegraphics[width=0.8\textwidth]{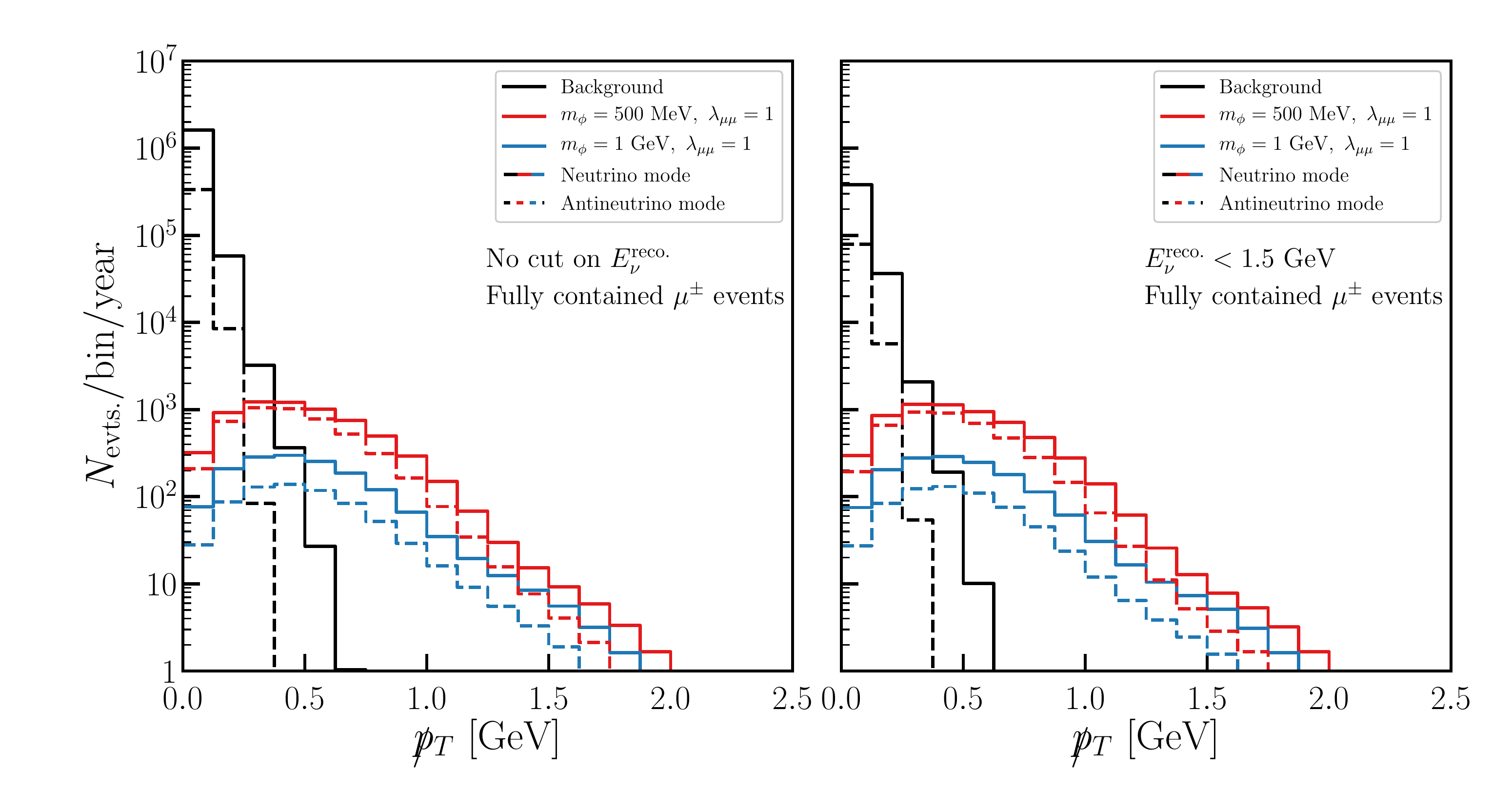}}
\caption{Expected number of events per year at the DUNE Near Detector that are identified as signal events in neutrino mode (solid lines) and antineutrino mode (dashed lines). See text and Tables~\ref{tab:NuMode} and~\ref{tab:NuBarMode} for explanation of signal. We show events for fully contained muons ($E_\mu < 1$ GeV), and in the right panel, we further restrict to events with reconstructed neutrino energy $E_\nu^\mathrm{reco} < 1.5$ GeV to highlight the difference in signal and background distributions as a function of this variable. For the mono-neutrino process of interest, red (blue) lines assume $m_\phi = 500$ MeV ($1$ GeV) and $\lambda_{\mu\mu} = 1$. Distributions take into account the factors regarding Michel electron tagging discussed in the text.}\label{METDist}
\end{figure}

Fig.~\ref{METDist} (left) shows the resulting $\cancel{p}_T$ distributions for our signal and background, using events with fully contained muons produced and then stopping inside the LAr TPC (with $E_\mu<1\,$GeV).\footnote[3]{Typically, a muon produced with energy larger than 1\,GeV will eventually leave the LAr TPC instead of stopping~\cite{MuonEnergyLoss}.
Hereafter, we use $E_\mu=1$\,GeV as the boundary between full contained muons within the LAr and the through-going muons that reach the GAr.
Based on the near detector design~\cite{DUNENDGAr}, the fraction of through-going muons that miss the GAr detector is neglected to a good approximation.}
The solid (dashed) curves correspond to the DUNE beam running in the neutrino (antineutrino) mode.
The neutrino mode has a relatively higher background because of its higher CC cross section (by a factor of 3).
For the signal, the red (blue) curves correspond to $m_\phi=500\,{\rm MeV}$ (1\,GeV) with the coupling parameter $\lambda_{\mu\mu}=1$ fixed.
Clearly, barring the uncertainties in the energy/momentum reconstruction, the signal events extend to much higher $\cancel{p}_T$ bins than the background.
The kinematic region with $\cancel{p}_T\gtrsim 0.5$\,GeV is expected to be almost background free given the time scale of the DUNE experiment ($\sim 10$\, years)
thus serves as the main signal region for hunting the mono-neutrino signal.

In addition, Ref.~\cite{Berryman:2018ogk} has pointed out that another observable, $E_\nu^\mathrm{reco}$, 
defined as the neutrino energy inferred by assuming $2\to 2$ scattering (where a neutrino with energy $E_\nu^\mathrm{reco}$ strikes a nucleon at rest), can be useful for further signal-background differentiation.
It was found that, for the SM background, $E_\nu^\mathrm{reco}$ peaks around 4\,GeV, whereas for our signal, $E_\nu^\mathrm{reco}$ is typically less than 1--2 GeV, mainly because the radiated $\phi$ carries away a significant fraction of the injected energy. 
In Fig.~\ref{METDist} (right), we have further imposed a $E_\nu^\mathrm{reco}<1.5\,$GeV cut in the event selection.

Even more excitingly, on top of the above kinematic handles\footnote{Another potentially useful kinematical cut is the invariant mass of the incoming neutrino, which can be inferred by assuming the target nucleon to be at rest and from the reconstructed final state momenta. Ideally, it vanishes for the background processes, but not for our signal if $\phi$ is radiated from the initial state with a nonzero transverse momentum. In practice, we find that the role of such a cut is already played by the $\cancel{p}_T$ cut.}
($\cancel{p}_T$ and $E_\nu^\mathrm{reco}$), the DUNE detectors will also be able to 
discriminate the signs of final state muons. The highly energetic muons ($E_\mu>1\,$GeV) will travel though the LAr and reach the magnetized HPg TPC (as shown by the yellow track ``c'' in Fig.~\ref{fig:schematic}). We simply assume 100\% charge discrimination efficiency in this case~\cite{DUNENDGAr}. 
The less energetic muons ($E_\mu<1\,$GeV) will mostly stop inside the LAr and become fully contained.
A $\mu^+$ always decays after stopping, giving rise to a Michel electron which can be tagged together with the stopped muon track (as shown by the yellow track ``a'' in Fig.~\ref{fig:schematic}).
In contrast, a $\mu^-$ is first captured by the argon nucleus into a muonic atom, and quickly settles down to the ground state~\cite{Vulpescu:1999vi}. Afterwards, it has $\sim 25\%$ probability of decaying, producing a Michel electron, and $\sim75\%$ probability of being further captured by the nucleus through weak interaction, converting into a muon neutrino, without producing Michel electron (as shown by the yellow track ``b'' in Fig.~\ref{fig:schematic}). These features can be used to further enhance the signal to background ratio.
For example, in the signal process (\ref{mononu}), a $\mu^+$ is produced which always decays into a Michel electron. For fully contained muon events, we can suppress the dominant background (\ref{Bkg1}) by a factor of 0.25 by requiring the presence of a Michel electron.
On the other hand, with an antineutrino beam, the dominant signal arises via the charge-conjugate process to (\ref{mononu}) and a $\mu^-$ is produced.
In this case, we could suppress the dominant background (the charge conjugation of (\ref{Bkg1}) which creates a $\mu^+$) almost completely by 
by vetoing events containing a Michel electron at the end of the stopped muon track, at the price of suppressing the signal by a factor of $0.75$ at the same time.
Recently, the MicroBooNE and ArgoNeuT experiments have demonstrated the ability of measuring Michel electrons and MeV scale physics in their LAr TPCs~\cite{Acciarri:2017sjy, Caratelli:2017chs, Acciarri:2018myr}.
A similar capability is expected at the DUNE near detector as well.

To better understand the signal and background event distributions in the different $\cancel{p}_T$ and $E_\nu^\mathrm{reco}$ kinematic regions, as well as the impact of the charge identification discussed above, we divide the generated events into several categories, 
as shown in the following Tables~\ref{tab:NuMode} and \ref{tab:NuBarMode}, assuming the DUNE beam running in the neutrino and antineutrino mode, respectively.
For each mode, we first divide the events into two cases with fully contained muons in LAr TPC (with $E_\mu < 1$\,GeV) and through-going muons that reach the HPg TPC (with $E_\mu >1$\,GeV). For each case, we further divide the events into two subsets with $E_\nu^\mathrm{reco}>(<)1.5$\,GeV.
In each $E_\mu$ and $E_\nu^\mathrm{reco}$ window, we further select events by applying the corresponding charge identifications. 
For events with muons fully contained in the LAr TPC, we require the presence (absence) of Michel electron near the end of the muon track, for the neutrino (antineutrino) beam mode. The resulting penalty factors for the signal and background are given in unit of percentage in the square brackets.
The HPg TPC is assume to veto all background events containing opposite-sign muons to those in the signal.

\begin{table}[h]
\begin{center}
\begin{tabular}{|c||c|c||c|c|}\hline
\multirow{2}{*}{$\nu$ \textbf{Mode}: $\mu^+$ Michel/Gas-tagged} & \multirow{2}{*}{$E_\mu < 1$ GeV} & $E_\nu^\mathrm{reco} < 1.5$ GeV & \multirow{2}{*}{$E_\mu > 1$ GeV} & $E_\nu^\mathrm{reco.} < 1.5$ GeV \\
& & $E_\nu^\mathrm{reco.} \geq 1.5$ GeV & & $E_\nu^\mathrm{reco} \geq 1.5$ GeV \\ \hline\hline
\multirow{2}{*}{{\rm Signal}: $\nu_\mu p \to \phi \mu^+ n$} & \multirow{2}{*}{Michel $e^+$ [100\%]} & $6.10\times10^3\ (1.53\times10^3)$ & \multirow{2}{*}{Tagged $\mu^+$ [100\%]} & $2.60\times10^3\ (660.4)$ \\
& & $423.4\ (59.9)$ & & $2.32\times10^3\ (516.4)$ \\ \hline
\multirow{2}{*}{{\rm Background}: $\nu_\mu n \to \mu^- p$} & \multirow{2}{*}{Michel $e^-$ [25\%]} & $3.71\times 10^5$ & \multirow{2}{*}{Tagged $\mu^-$ [0\%]} & $5.27 \times 10^5$ \\
& & $1.23\times 10^6$ & & $1.55\times 10^7$ \\ \hline
\multirow{2}{*}{{\rm Background}: $\overline{\nu}_\mu p \to \mu^+ n$} & \multirow{2}{*}{Michel $e^+$ [100\%]} & $4.33\times 10^4$ & \multirow{2}{*}{Tagged $\mu^+$ [100\%]} & $3.41\times 10^4$ \\
& & $2.28\times 10^4$ & & $6.30\times 10^5$ \\ \hline
\end{tabular}
\caption{Number of expected signal and background events per year at the DUNE near detector, assumed to contain 75 ton of liquid argon, with the beam running in the neutrino mode. 
We divide the events into four kinematical categories with $E_\mu < (>) 1$ GeV and $E_\nu^\mathrm{reco} > (<) 1.5$\,GeV.
For charge identification, we require the presence of Michel electrons from fully contained muons inside the LAr TPC or tagged $\mu^+$ inside the HPg TPC.
The resulting penalty factors for the signal and background are given in unit of percentage in the square brackets.
We show the number of signal events for $\phi$ emission assuming $\lambda_{\mu\mu} = 1$ and $m_\phi = 500$ MeV ($m_\phi = 1$ GeV in the parentheses).}\label{tab:NuMode}
\end{center}
\end{table}

\begin{table}[h]
\begin{center}
\begin{tabular}{|c||c|c||c|c|}\hline
\multirow{2}{*}{$\bar{\nu}$ \textbf{Mode}: $\mu^-$ No Michel/Gas-tagged} & \multirow{2}{*}{$E_\mu < 1$ GeV} & $E_\nu^\mathrm{reco} < 1.5$ GeV & \multirow{2}{*}{$E_\mu \geq 1$ GeV} & $E_\nu^\mathrm{reco.} < 1.5$ GeV \\
& & $E_\nu^\mathrm{reco.} \geq 1.5$ GeV & & $E_\nu^\mathrm{reco} \geq 1.5$ GeV \\ \hline\hline
\multirow{2}{*}{{\rm Signal}: $\bar{\nu}_\mu n \to \phi \mu^- p$} & \multirow{2}{*}{Nuclear captured [75\%]} & $3.30\times10^3 (486.2)$ & \multirow{2}{*}{Tagged $\mu^-$ [100\%]} & $1.46\times10^3 (216.7)$ \\
& & $409.5 (40.8)$ & & $1.67\times10^3 (222.3)$ \\ \hline
\multirow{2}{*}{{\rm Background}: $\overline{\nu}_\mu p \to \mu^+ n$} & \multirow{2}{*}{Michel $e^+$ [0\%]} & $5.38\times 10^5$ & \multirow{2}{*}{Tagged $\mu^+$ [0\%]} & $6.30\times 10^5$ \\
& & $4.43\times 10^5$ & & $6.18\times 10^6$ \\ \hline
\multirow{2}{*}{{\rm Background}: $\nu_\mu n \to \mu^- p$} & \multirow{2}{*}{Nuclear captured [75\%]} & $6.34\times 10^4$ & \multirow{2}{*}{Tagged $\mu^-$ [100\%]} & $3.30\times 10^4$ \\
& & $1.95\times 10^5$ & & $1.94\times 10^6$ \\ \hline
\end{tabular}
\caption{Similar to Table~\ref{tab:NuMode} but with the DUNE beam running in the antineutrino mode.
For charge identification, we require the absence of any Michel electron for fully contained muons inside the LAr TPC or tagged $\mu^-$ inside the HPg TPC in this case.}
\label{tab:NuBarMode}
\end{center}
\end{table}

From these tables, there is important information one could learn about the features of the signal and background.
First, the dominant background category contains through-going muons that reach the HPg TPC.
Among them, the opposite-sign muon background (compared to the signal) can be vetoed with a magnetic field.
Second, the majority of signal events have fully contained muons in the LAr TPC.
The number of background events can be suppressed by an order of magnitude by focusing on these fully contained muons.
Furthermore, we find that the background can be suppressed by another order of magnitude by requiring the presence/absence of a Michel electron from muon decay together with a $E_\nu^\mathrm{reco}<1.5$\,GeV cut. 
In contrast, these event selection criteria only hurt the number of signal events by an order one factor.

\subsection{Expected DUNE reach}

Here we assess the prospects of using the DUNE near detectors to explore the mono-neutrino signal, and their implications for the neutrinophilic dark matter models presented in section~\ref{sec:benchmark}.
In order to fully take advantage of the differences between the signal and background as found above, we perform a log-likelihood comparison between the two using a Poisson distribution
\begin{equation}
-2\ln{\mathcal{L}} = -2 \sum_{C,T} \sum_{i=1}^{24} \sum_{j=1}^{20} \left( -\lambda_{ij} + \mu_{ij} \ln{\lambda_{ij}} - \ln{\mu_{ij}!} \right),
\end{equation}
where $\mu_{ij}$ ($\lambda_{ij}$) is the number of background (signal + background) events~\footnote[4]{The difference between $\lambda_{ij}$ and $\mu_{ij}$ -- the number of signal events in a bin -- is proportional to $\lambda_{\mu\mu}^2$.} in the $i$th bin of $\cancel{p}_T$ and $j$th bin of $E^\mathrm{reco}_\nu$. We assume ten years of data collection (equal time in neutrino and antineutrino modes) and bins of $125$ MeV in $\cancel{p}_T$ between $0$ and $3$ GeV (24 total) and bins of $500$ MeV in $E_\nu^\mathrm{reco}$ between $0$ and $10$ GeV (20 total).

We perform the following two analyses. The first one involves both fully contained ($C$) events within the LAr TPC and the ones with through-going events ($T$) muons that could reach the HPg TPC. We take into account the charge identification efficiency factors discussed above. The second analysis involves only the $C$-type events, and to be conservative, in this case we explore the impact of having {\it no} charge identification, meaning no discrimination between captured $\mu^-$ and those that produce a Michel electron/positron after stopping.
This amounts to the factors in brackets in Tables~\ref{tab:NuMode} and~\ref{tab:NuBarMode} all being $100\%$ -- an increase in background by a factor of $\sim 4$ in neutrino mode and a factor of a few in antineutrino mode.

In Fig.~\ref{moneyplot}, the black curves show the 90\% confidence level reach in the $\lambda_{\mu\mu}$ versus $m_\phi$ parameter space (corresponding to $-2\Delta (\ln{\mathcal{L}}) = 4.6$ for two parameters) by using the DUNE near detector to look for the proposed mono-neutrino signal for 10 years.
The solid (dot-dashed) curve corresponds to using both $C$ and $T$ (just $C$) events.
As explained earlier, the colored curves represent the parameter space for neutrinophilic dark matter which obtains the correct thermal relic abundance via the annihilation into neutrinos. 
Clearly, a sizable portion of these regions of interest can be covered.
Including the $T$-type events and enabling the charge identification capability of detectors allows the sensitivity in $\lambda_{\mu\mu}$ to be enhanced by a factor of $\sim2$.

\begin{figure}[t]
\centerline{\includegraphics[width=0.8\textwidth]{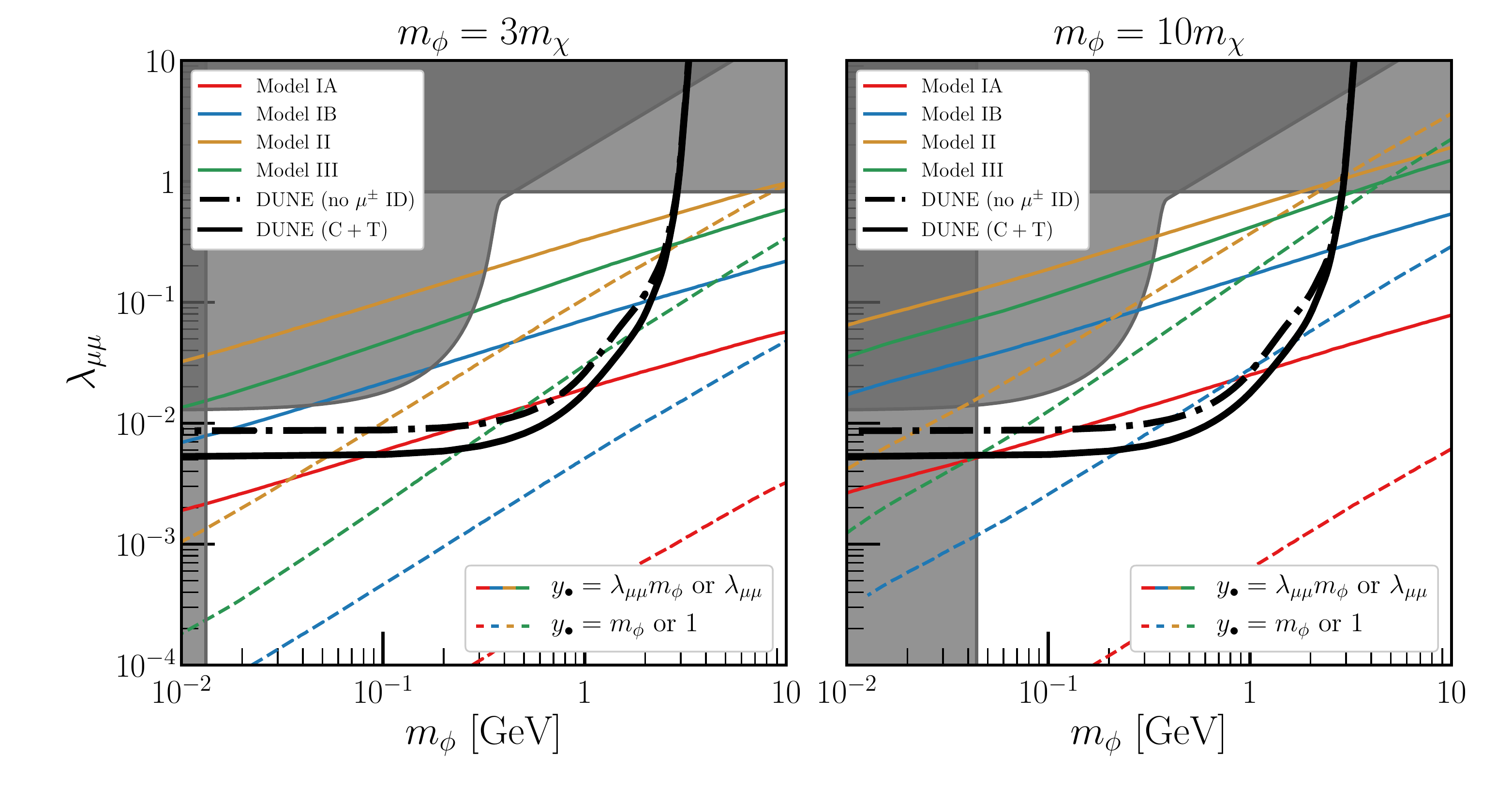}}
\caption{Left: Same as Fig.~\ref{moneyplot} but assuming a mass relation $m_\phi = 3m_\chi$.
Right: a reproduction of Fig.~\ref{moneyplot} for comparison. }\label{fig:RDTwo}
\end{figure}

In our discussions so far, we have sticked the $\phi$, $\chi$ mass relation to be $m_\phi = 10m_\chi$. To present our results for generalized mass relations,
In the left panel of Fig.~\ref{fig:RDTwo}, we show the parameter space of thermal targets versus the DUNE reach assuming $m_\phi = 3m_\chi$.
In this case, most of the relic density curves for the neutrinophilic dark matter models discussed in Section~\ref{sec:benchmark} move downwards slightly, compared Fig.~\ref{moneyplot} (shown again in the right panel). Clearly, the DUNE is still able to cover a substantial portion of the parameter space of interest.
On the other hand, if we increase the $m_\phi/m_\chi$ mass ratio, the correct dark matter relic density will call for larger couplings and the DUNE reach will get better.

\section{Other constraints}\label{sec:otherbounds}

In this section, we comment on other existing constraints on the benchmark models under consideration.

\subsection{Precision Measurements in the Standard Model}

There are relevant existing constraints on the $\phi$ couplings from precision measurements in the SM, as shown by the gray regions in Fig.~\ref{moneyplot}.
The leading ones are the charged meson ($\mathfrak{m} = \pi, K, D$) decays and the Higgs boson invisible decays.
In particular, the semi-leptonic decay of charged kaon, $K^-\to \ell^- + \bar\nu_\ell$, is helicity suppressed by the mass of the charged lepton $\ell$, which will no longer be the case if a $\phi$ particle is radiated off the final state neutrino~\cite{Barger:1981vd, Laha:2013xua}. 
The three-body decay meson rate for $\mathfrak{m}^-\to \ell + \nu_\ell + \phi$ has been calculated in Ref.~\cite{Berryman:2018ogk}, which is valid when $m_\phi < m_\mathfrak{m} - m_\ell$. In this work, we extend the result to the $m_\phi > m_\mathfrak{m} - m_\ell$ region and calculate the four-body decay rate $\mathfrak{m}^-\to \ell^- + \nu_\ell + \bar\nu + \bar\nu$ via an off-shell $\phi$ in Appendix~\ref{app:4body}.
The kaon decay constraint excludes the gray shaded region in the upper-left corner of Figs.~\ref{moneyplot} and~\ref{fig:RDTwo}.

The effective operator in Eq.~(\ref{Leff}) also leads to a new Higgs boson decay mode, $h\to \phi \nu\nu$. 
The upper bound on the Higgs invisible decay width from the LHC translates into an upper limit on $\lambda_{\mu\mu}$~\cite{Berryman:2018ogk}, shown by the gray shaded region near the top of Figs.~\ref{moneyplot} and~\ref{fig:RDTwo}. 
The $\phi$ radiation also modifies the $W/Z$ boson decay widths which lead to weaker constraints than the ones discussed above.

\subsection{Lower Bound on Thermal Dark Matter Mass from CMB}

For the dark matter to obtain its relic density via the thermal freeze out mechanism, there is a generic lower bound on its mass from the CMB measurement of the effective number of relativistic degrees of freedom, $\Delta N_\mathrm{eff}$~\cite{Aghanim:2018eyx}. If the dark matter is too light, the transfer of entropy into its annihilation product will increase $\Delta N_\mathrm{eff}$; this leads to a lower bound on dark matter mass around 4\,MeV~\cite{Nollett:2014lwa, Escudero:2018mvt}. We include this lower bound, translated into a lower bound on $m_\phi$ depending on the $\phi$-$\chi$ mass ratio, in our main results in Figs.~\ref{moneyplot} and~\ref{fig:RDTwo}.
Recently, it has been argued that if dark matter enters thermal equilibrium with neutrinos after the BBN, the CMB constraint can be relaxed~\cite{Berlin:2017ftj}.
However, the dark-matter-neutrino coupling for this to occur must be very small, and cannot lead to a large enough mono-neutrino rate at DUNE.

\subsection{Constraint on Dark Matter Self Interaction}

The $\phi$ exchange can also give rise to dark matter self interactions among the dark matter particles, whose cross section is bounded from above from the observation of the bullet cluster~\cite{Randall:2007ph}, ${\sigma_{\chi\chi \to \chi\chi}}/{m_\chi} \lesssim 1.25\,{\rm cm^2}/{\rm g} \simeq 4.6\times 10^3\,{\rm GeV}^{-3}$.
Here we discuss the implication of this constraint in each model.

\smallskip
{\it Model IA}: For each $\chi$ particle, there are two ways for it to self interact, $\chi \chi\to \chi\chi$ via $s$-channel $\phi$ exchange, and $\chi \bar\chi\to \chi \bar\chi$ via $t$-channel $\phi$ exchange. In the large $m_\phi$ limit, the two cross sections are comparable to each other,
\begin{equation}
\sigma_{\chi\chi\to\chi\chi} = \frac{1}{2} \sigma_{\chi\bar\chi\to\chi\bar\chi} = \frac{y_{IA}^4}{128\pi m_\chi^2 m_\phi^4} \ .
\end{equation}

\smallskip
{\it Model IB}: The channels for $\chi$ to self interact are like model IA. The cross sections in this case are
\begin{equation}
\sigma_{\chi\chi\to\chi\chi} = \frac{y_{IB}^4 m_\chi^2 v_{\rm rel}^4}{128\pi m_\phi^4} \ ,  \ \ \ 
\sigma_{\chi\bar\chi\to\chi\bar\chi} = \frac{y_{IB}^4 m_\chi^2}{4\pi m_\phi^4} \ .
\end{equation}
Interestingly, the $\chi\chi\to\chi\chi$ process (via $s$-channel $\phi$ exchange) is highly suppressed at low velocities, which is the case for dark matter on cluster scales ($v_{\rm rel} \sim 0.03c$). This implies that if the dark matter relic abundance is asymmetric today, made of only $\chi$ or $\bar\chi$ particles, its self interaction constraint is more relaxed.

\smallskip
{\it Model II, III}: In these two models, the dark matter self interactions occur at one loop level. 
The cross section in model II is
\begin{equation}
\sigma_{\chi \chi\to \chi \chi} \sim \sigma_{\chi \bar\chi\to \chi \bar\chi} \simeq \frac{1}{64\pi m_\chi^2} \left[\frac{y_{II}^2}{16\pi^2} \left( \ln \frac{\mu^2}{m_\phi^2} - \frac{m_\chi^2}{m_\phi^2 - m_\chi^2} \ln \frac{m_\phi^2}{m_\chi^2} \right)\right]^2 \ ,
\end{equation}
where the logarithmic divergence indicates that the $|\chi|^4$ coupling is renormalized by the $\chi^3\phi$ coupling at one-loop level, and the $\chi$ self interaction is not calculable. In practice, we get rid of the cutoff scale dependence by assuming the bare $|\chi|^4$ coupling vanishes at the scale $\mu=m_\phi$.
 We estimate the cross sections in model III as
\begin{equation}
\sigma_{\chi_1 {\chi}_1 \to \chi_1 {\chi}_1} \sim \sigma_{\chi_1 \bar{\chi}_1 \to \chi_1 \bar{\chi}_1} \simeq \frac{y_{III}^8 m_\chi^2}{16 \pi m_\phi^4} \left(\frac{1}{16\pi^2}\right)^2 \ .
\end{equation}

In each model, we explore the implications of self interaction constraint, $\sigma/m_\chi \lesssim 1.25$ cm$^2$/g, on its parameter space. 
To present the results, we fix $m_\chi = m_\phi/10$ and consider the same $\phi$-$\chi$ couplings as used in Fig.~\ref{moneyplot}.
For model IA, with $y_{IA}/m_\phi=1$, small enough self interaction requires $m_\phi \gtrsim 106$ MeV.
For the other models, with $y_\bullet = 1$ ($\bullet=IB, II, III$), we derive lower bounds $m_\phi \gtrsim 12$ MeV, $6$ MeV, $0.2$ MeV, respectively. 
These lower bounds, when translated into bounds on $m_\chi$, are comparable to those from the CMB discussed in the previous subsection.
On the other hand, assuming $y_{IA}/m_\phi = \lambda_{\mu\mu}$ or $y_\bullet = \lambda_{\mu\mu}$ in the other models, the self interaction constraints on the parameter space are much weaker than the other existing bounds.

Because we have assumed the force carrier $\phi$ to be heavier than the dark matter in this work, the above self interaction strength can be safely suppressed in the large $m_\phi$ limit. 
Within the parameter space of interest to this study, we find it safe to conclude that the dark matter self interaction constraint is always satisfied. 
In the complementary regime, making $\phi$ lighter than the dark matter could result in resonant enhancements to the self interaction cross section at low dark matter velocities~\cite{Buckley:2009in, Tulin:2012wi} and the constraint from clusters is typically much more severe. On the other hand, it is possible to accommodate a self-interacting dark matter candidate in this case~\cite{Rajaraman:2018bam, Tulin:2017ara}.

\subsection{IceCube and High-energy Cosmic Neutrinos Interacting with Dark Matter}

In the models we consider, the $\phi$ exchange gives rise to neutrino self interactions as well as neutrino-dark-matter interactions. The observation of very high energy cosmic neutrinos at IceCube constrains these interactions because they have to survive traveling through the ambient cosmic neutrino and dark matter backgrounds before reaching the earth~\cite{Ng:2014pca, Cherry:2014xra}. Moreover, the coincidence in the recent observations of high energy neutrinos and gamma rays indicates the distance of the source, allowing one to use their relative flux to limit the high energy neutrino free streaming length~\cite{Kelly:2018tyg}. We examine this constraint and find that, for dark matter and $\phi$ heavier than MeV scale, the focus of this work, the IceCube constraint is weaker than other existing ones.

\section{Summary and Outlook}\label{sec:conclusion}

In this work, we present the {\it mono-neutrino} process as a new signal and powerful probe of neutrinophilic dark matter candidates, that can be searched for at neutrino experiments. The main idea is to consider the radiation of light invisible particle(s) off the neutrino beam when a charge-current weak process occurs for neutrino to get detected. This leads to a missing transverse momentum in the final states with respect to the incoming neutrino beam direction, which is a clean signal with low standard model background.

We focus on a class of models where a complex scalar $\phi$ with lepton number $-2$ interacts with standard model neutrinos via the $(LH)^2\phi$ operator. This way, the Higgs condensate, by breaking the $SU(2)_L$ gauge invariance, allows $\phi$ to couple exclusively to neutrinos. We discuss several options for $\phi$ to be the portal to dark matter, where the latter is a standard model singlet and is stabilized due to various global symmetries, $\mathbb{Z}_2$, $\mathbb{Z}_3$ and $U(1)$. We derive the model parameter space that allows the dark matter to obtain the correct thermal relic abundance by annihilating into neutrinos through $\phi$. The results serve as well motivated targets for the mono-neutrino search proposed here. 
Moreover, the radiation of $\phi$ from the neutrino beam carries away two units of lepton number. Thus, it also leads to ``wrong sign'' charged lepton production -- another striking signal that accompanies the missing transvers momentum.

We investigate the feasibility and prospects of using the DUNE near detectors to hunt for the above signals, including the liquid argon TPC, for final state tracking and momentum reconstruction, and a prospective magnetized gaseous argon TPC, which is good at lepton charge identification.
We estimate the standard model background based on a simplified nucleon level simulation with a final momentum smearing.
Based on our analysis, we find it is well motivated and feasible for the DUNE, as a multipurpose experiment, to efficiently probe the presently allowed parameter space of neutrinophilic thermal dark matter.
Our result could be further improved in the future with more precise modeling of nuclear recoils (see~\cite{Friedland:2018vry} for a recent discussion on this topic) as well as realistic detector simulations.
The search strategy discussed here is applicable to other neutrino experiments such as the short baseline neutrino program at Fermilab.

Although our discussion of the mono-neutrino signal has focused on the coupling of $\phi$ to muon neutrinos, we hope it could inspire future considerations on other neutrino flavors by exploring a broader range of neutrino sources and detectors.

\begin{acknowledgements}
We thank Jeffrey Berryman, Zackaria Chacko, Andr\'e de Gouv\^ea, Roni Harnik, Joshua Isaacson, Chris Marshall and Hai-Bo Yu for useful discussions.
This manuscript has been authored by Fermi Research Alliance, LLC under Contract No. DE-AC02-07CH11359 with the U.S. Department of Energy, Office of Science, Office of High Energy Physics. 
The work of Y.Z. is also supported by the DoE under contract number DE-SC0007859.
Y.Z. acknowledges a travel support from the Colegio De Fisica Fundamental E Interdisciplinaria De Las Americas (COFI).
\end{acknowledgements}

\appendix
\section{Four Body Charged Meson Decay Through Off-shell Mediator $\bg{\phi}$}\label{app:4body}

Here we discuss the four-body final state of a meson decay, $\mathfrak{m^+} \to \ell^+ \overline{\nu} \phi^*,$ $\phi^* \to \nu\nu$ (where $\phi^*$ denotes that the $\phi$ is off-shell and can even be heavier than the meson $\mathfrak{m}$ in this case) and its charge-conjugate. Following the four-body phase space developed in Ref.~\cite{Anastasiou:2003gr} for massless final-state particles and extended to include one massive particle (in our case, $\ell$) in Ref.~\cite{Asatrian:2012tp}, we can write the width of this process as
\begin{equation}
\Gamma_{\mathfrak{m^\pm}\to\ell^\pm \nu\nu\overline{\nu}} = \frac{\lambda_{\alpha\beta}^4 G_F^2 f_\mathfrak{m}^2 m_\mathfrak{m}^5}{2^{10} \pi^{6}} \int g(\vec{\lambda}; m_\mathfrak{m}, m_\phi, m_\ell) d^{5}\vec{\lambda} \ ,
\end{equation}
where $G_F$ is the Fermi weak coupling; $f_\mathfrak{m}$ is the decay constant of meson $\mathfrak{m}$; $m_\mathfrak{m}$, $m_\phi$, and $m_\ell$ are the masses of $\mathfrak{m}$, the mononeutrino $\phi$, and charged lepton $\ell$, respectively, and $g(\vec{\lambda}; m_\mathfrak{m}, m_\phi, m_\ell)$ is a function of the parameters $\lambda_i$, $i = 1, \dots, 5$. The integration over $d^5\vec{\lambda}$ ranges over $\lambda_i = 0$ to $\lambda_i = 1$ for all $i$. We write $g(\vec{\lambda}; m_\mathfrak{m}, m_\phi, m_\ell)$ as
\begin{equation}
\begin{split}
g(\vec{\lambda}; m_\mathfrak{m}, m_\phi, m_\ell) &= \frac{(1-\lambda_2)^2 \lambda_2 m_\mathfrak{m}^2 (1-\sqrt{x_1})^4 s_{\lambda_1}}{4\sqrt{\lambda_5(1-\lambda_5)} \left(\lambda_1(1-\lambda_2)\lambda_4 m_\mathfrak{m}^2 m_\phi^2 \left( f_\Gamma^2 - 2(1-\sqrt{x_1})^2\right) + \lambda_1^2 (1-\lambda_2)^2 \lambda_4^2 m_\mathfrak{m}^4(1-\sqrt{x_1})^4 + m_\phi^4\right)}  \\
&\times\left[ (\lambda_1-1)(2\lambda_3-1) s_{\lambda_1} \left(\lambda_1(1-\sqrt{x_1})^2+x_1-1\right) + \rule{0mm}{5mm}\right.\\
&\left. \sqrt{\lambda_1-1}\left((2\lambda_3-1)(1+\sqrt{x_1})\sqrt{\lambda_1^2 - 2\lambda_1(1+x_1)+(1-x_1)^2}+\lambda_1^2(\sqrt{x_1}-1)^3+\lambda_1(1+\sqrt{\lambda_1})\right)\right] \ ,
\end{split}
\end{equation}
where $s_{\lambda_1} \equiv \sqrt{\lambda_1 + (\lambda_1-1)x_1 - 2(1+\lambda_1)\sqrt{x_1}-1}$. The integrals over $\lambda_3$ and $\lambda_5$ are straightforward, and we arrive at $h(\vec{\lambda}; m_\mathfrak{m}, m_\phi, m_\ell) \equiv \int g(\vec{\lambda}; m_\mathfrak{m}, m_\phi, m_\ell) d\lambda_3 d\lambda_5$,
\begin{equation}
h(\vec{\lambda}; m_\mathfrak{m}, m_\phi, m_\ell) = \frac{\pi m_\mathfrak{m}^2 \lambda_1 \lambda_2 (1-\lambda_2)^2 (1-\sqrt{x_1})^4 (1 + \sqrt{x_1} + \lambda_1(\sqrt{x_1}-1)^3) s_{\lambda_1}}{4 \left(\lambda_1 (1-\lambda_2) \lambda_4 (f_\Gamma^2 - 2(1-\sqrt{x_1})^2) m_\mathfrak{m}^2 m_\phi^2 + \lambda_1^2 (1-\lambda_2)^2 \lambda_4^2 (1-\sqrt{x_1})^4 m_\mathfrak{m}^4 + m_\phi^4\right)} \ .
\end{equation}

For large $m_\phi \gg m_\mathfrak{m}$, this decay width can be solved analytically. Keeping terms to $\mathcal{O}(m_\mu/m_\mathfrak{m})$, we have the result:
\begin{equation}
\Gamma_{\mathfrak{m^\pm}\to\ell^\pm \nu\nu\overline{\nu}} = \frac{\lambda_{\alpha\beta}^4 G_F^2 f_\mathfrak{m}^2 m_\mathfrak{m}^6 (m_\mathfrak{m}+4m_\mu)}{2^{16} 3^2\pi^{5} m_\phi^4} \ .
\end{equation}

\bibliographystyle{apsrev-title}
\bibliography{MonoReferences}{}

\end{document}